\documentclass{ws-procs975x65}

\begin{document}

\title{``SINGULARITIES" IN SPACETIMES WITH DIVERGING HIGHER-ORDER CURVATURE INVARIANTS}

\author{D. A. KONKOWSKI}

\address{Department of Mathematics, U.S. Naval Academy\\
Annapolis, Maryland, 21012, USA\\
E-mail: dak@usna.edu}

\author{T.M. HELLIWELL}

\address{Department of Physics, Harvey Mudd College\\
Claremont, California, 91711, USA\\
E-mail: helliwell@HMC.edu}

\begin{abstract}
After reviewing the definitions of classical and quantum singularities, it is shown by example that if zeroth-order curvature invariants are regular, a diverging higher-order curvature invariant does not necessarily imply the existence of a classical or a quantum singularity.
\end{abstract}

\bodymatter

\section{Introduction}
A spacetime $(M, g)$ is a smooth, $C^\infty$, paracompact, connected Hausdorff manifold $M$ with a Lorentzian metric $g$. Here we present three spacetimes \cite{Bonnor, ML, Siklos} with regular zeroth-order curvature invariants but diverging higher-order invariants and illustrate with one spacetime \cite{ML} (the Musgrave-Lake ST) that such a divergence does not necessarily foretell the existence of a ``singularity" using the usual definitions. 

\section{Singularity Definitions}

\subsection{Classical Singularities}
A classical singularity is indicated by incomplete geodesics or incomplete paths of bounded acceleration \cite{HE, Geroch} in a maximal spacetime. Since, by definition, a spacetime is smooth, all irregular points (singularities) have been excised; a singular point is a boundary point of the spacetime. There are 3 types of singularities\cite{ES}: quasi-regular (a mild, topological type), non-scalar curvature (diverging tidal forces on curves ending at the singularity; finite tidal forces on some nearby curves) and scalar curvature (diverging scalars -- usually only consider $C^0$ scalar polynomial (s.p.) invariants). 

\subsection{Quantum Singularities}
A spacetime is QM (quantum-mechanically) nonsingular if the evolution of a test scalar wave packet, representing the quantum particle, is uniquely determined by the initial wave packet, manifold and metric, without having to put boundary conditions at the singularity\cite{HM}. Technically, a static ST is QM-singular if the spatial portion of the relevant wave operator, here the Klein-Gordon operator, is not essentially self-adjoint\cite{RS} on $C_{0}^{\infty}(\Sigma)$ in $L^2(\Sigma)$ where $\Sigma$ is a spatial slice.

\section{Spacetimes with Diverging Higher-Order Curvature Invariants}
As Musgrave and Lake say, ``curvature invariants alone are not sufficient to probe the `physics' of the solution \cite{ML}."

\subsection{Kinnersley 'photon rocket'}
Bonnor \cite{Bonnor} analyzed the Kinnersley\cite{Ki} `photon rocket' which has two-metric functions, the mass $m = m(u)$ and the acceleration $a = a(u)$, both functions of the radial null coordinate $u$. He found that $a(u)$ does not enter any zeroth-order s.p. curvature scalars, but it does enter into differential invariants. Thus, a singular acceleration `singularity' would not show up on regular curvature invariants but are crucial for an adequate physical picture and predict a true physical singularity since they indicate incomplete, inextendible null geodesics.

\subsection{Siklos Whimper Spacetimes}
Siklos\cite{Siklos} in 1976 considered the so-called ``whimper" STs (``not with a bang, but with a whimper" as the poet T.S. Elliot wrote \cite{Elliot}). These STs are geodesically incomplete and inextendible and thus classically singular. They possess $C^0$ non-scalar curvature singularities; all zeroth-order s.p. invariants are regular. However, Siklos did find these STs to have diverging  first-order curvature invariants. 

\subsection{Musgrave-Lake Spacetimes}
Musgrave-Lake STs \cite{ML} are static and spherically-symmetric with metric 

\begin{equation}
ds^2 = -(1 + r^{n+3/2}) dt^2 + (1 + r^{n+3/2}) dr^2 + r^2 d\theta^2 + r^2 \sin^2(\theta) d\phi^2
\end{equation}

\noindent where the coordinates have their usual ranges and the parameter $n$ = 1,2,3,4. They have an anisotropic matter distribution and obey all energy conditions (weak, strong, dominate). Physically, they can be interpreted as a ``thick shell" with a density and pressure that approach zero as $r\rightarrow0$ or as $r\rightarrow\infty$.  All $C^0$ s.p. curvature invariants vanish at $r=0$. Differential invariants up to order $(n-1)$ are regular at $r=0$ while $n^{th}$ order differential invariants diverge at $r=0$. However, there are no incomplete geodesics and hence no classical singularity \footnote{ Actually, for the $n=1$ case, the tangential pressure is not $C^1$ and this can be considered the physical reason why the first-order differential invariants diverge\cite{ML}.} in the usual sense. Observers following timelike and null geodesic paths feel nothing untoward at $r=0$. In particular, tidal forces do not diverge. \smallskip

\noindent The quantum singularity structure of the Musgrave-Lake spacetime was also studied. The massive Klein-Gordon equation was solved, variables separated, and radial solutions approximated near $r=0$. Both radial solutions are square integrable, but one does not form a proper solution of the three dimensional wave equation for the spatial wave function. This can be seen by combining each radial solution with the $l=0, m=0$ angular spherical harmonic, intergrating over a spherical volume, and finding a nonzero value for one integral. Thus, the Musgrave-Lake spacetime is quantum mechanically non-singular.

\section{Conclusions}
Spacetimes with higher-order diverging invariants have interesting "singularity" structure. Further study of these and other cases is warrented.

\section{Acknowledgments}
One of us (DAK) thanks Professor Malcolm MacCallum for useful discussions and Queen Mary, University of London, where some of this work was carried out.

\bibliographystyle{ws-procs975x65}
\bibliography{ws-pro-sample}

\end{document}